\documentclass[12pt]{article}
\usepackage{natbib}
\usepackage{amsmath}
\usepackage[margin=1.2in]{geometry}

\title{Inter-theory Relations in Quantum Gravity: Correspondence, Reduction, and Emergence}
\author{Karen Crowther\\\textit{Department of Philosophy, University of Geneva}}
\date{} 

\begin{document}
	\maketitle
	\bibliographystyle{newapa}
	
	\begin{abstract}
Relationships between current theories, and relationships between current theories and the sought theory of quantum gravity (QG), play an essential role in motivating the need for QG, aiding the search for QG, and defining what would count as QG. \textit{Correspondence} is the broad class of inter-theory relationships intended to demonstrate the necessary compatibility of two theories whose domains of validity overlap, in the overlap regions. The variety of roles that correspondence plays in the search for QG are illustrated, using examples from specific QG approaches. \textit{Reduction} is argued to be a special case of correspondence, and to form part of the definition of QG. Finally, the appropriate account of \textit{emergence} in the context of QG is presented, and compared to conceptions of emergence in the broader philosophy literature. It is argued that, while emergence is likely to hold between QG and general relativity, emergence is not part of the definition of QG, and nor can it serve usefully in the development and justification of the new theory.


	\end{abstract}

\tableofcontents
	
\section{Introduction}\label{Intro}

The search for a theory of quantum gravity (QG) is one of the biggest open problems in physics.\footnote{``QG'' can refer to either the collective research program, or the theory being sought. Here, it is usually the \textit{theory} that is being referred to; I endeavour to be explicit when context demands it.} The theory must describe the domains where both general relativity (GR) and quantum (field) theory are supposed to be required. It is generally expected to replace GR, in the sense of being a \textit{more fundamental} theory that describes the ``high-energy'' or ``micro'' degrees of freedom that ``underlie'' gravitational phenomena.\footnote{\label{microfoot}Here, ``micro'' is used purely in a figurative sense, as a means of distinguishing the degrees of freedom described by QG from those (``macro'' degrees of freedom) of current physics. ``High-energy scales'' and ``short-length scales'' are used interchangeably, and are also used to signify the domain expected to be described by QG. The scare quotes indicate that this may not literally be true, because the idea of length (and, correspondingly, energy) may cease to be meaningful at some scale, and QG may describe this very regime. Please keep this in mind, as I will drop the scare quotes.} Exploring the relationships between QG and current physical theories is interesting in itself, but is important also for myriad other reasons, including better understanding the nature and implications of current theories; investigating scientific theory-change and theory construction; and gaining insight into the nature and structure of QG. In fact, understanding these relationships serves crucially in the development and justification of QG. In regards to theory development, for instance, the links between current theories play a role in indicating the domain where QG is necessary, and act as stepping-stones towards finding the new theory. And, in regards to theory justification, the requirement that the new theory link back to current theories is especially significant in the case of QG, given the lack of novel empirical data that QG is required to explain, and the extreme regimes where QG is expected to be needed---it seems like this link back to established physics may be the surest way (currently known) for QG to make contact with the empirical realm.

\paragraph*{}
Additionally, there is an even stronger sense in which understanding these relationships is crucial to theory development and justification in QG---they play a role in determining what would count as a theory, i.e., in \textit{defining what QG is}. Here, QG is understood as any theory that satisfies the set of criteria that are taken to define QG. Currently, there is no well-established, generally agreed-upon set; however, some of the criteria whose \textit{inclusion} is the least controversial across all the approaches to QG concern the relationships to, and between, current theories. For instance, I take it that the set includes at least the following criteria: that the theory describe the domains where both GR and quantum theory are necessary\footnote{Note that gravity is a universal force, and, technically, quantum theory is also universal: its domain of applicability includes all physical systems. Yet, we do not \textit{need} to use GR and quantum theory to describe all systems.}; that the theory ``recover'' GR in the regimes where GR is known to be successful; and that the theory take into account quantum theory (by being quantum itself, in some sense, or by also relating to the framework of quantum field theory, or recovering particular quantum field theories---or a combination of these methods).\footnote{I have been purposely vague in framing each of these criteria, because even though this list represents part of the minimal agreed upon set, the \textit{interpretations} of these criteria vary from approach to approach.}

\paragraph*{}
As I show in this paper, the inter-theory relationships that serve in the roles concerning the development and justification of QG (including, for instance, expounding and refining the relevant notion of ``recovery'' that features in the definition of QG, and demonstrating the relative fundamentality of the theory) are those that are broadly classifiable as \textit{correspondence} and \textit{reduction}.\footnote{Some of the various other ways in which these relationships are used in the search for QG are outlined and illustrated in the relevant sections below (\S\ref{Corresp}--\ref{RedQG}).}  Yet, so far, the literature on the philosophy of QG has not looked at these relationships. Instead, however, there has been a lot of work considering the idea of \textit{emergence} in the context of QG. This literature typically makes little attempt to explicate its connection to the main accounts of emergence in more general literature,\footnote{There are exceptions, e.g., \citet{Bain2013b}.} and often either works with a deliberately minimal characterisation of ``emergence'', as a generic asymmetric inter-theory relation (involving one theory being more fundamental than another)\footnote{This is not a criticism of the individual papers, however, since the minimal characterisation may be sufficient for these authors' arguments, e.g., \citet{Knox2013, Teh2013, Rickles2013, Seiberg2007}.}, or construes ``emergence'' as something akin to (but perhaps running in the opposite direction to) reduction---being a relation, or relations, illustrating (broadly) the dependence, derivability, or ``recovery'' of GR (or aspects/structures of GR, including spacetime) from QG.\footnote{E.g., \citet{Berenstein2006, Butterfield1999, Wuthrich2017, Yang2009}.} The best way of understanding emergence in the context of QG, however, is as a relationship where a less-fundamental theory is novel and autonomous (or ``robust'') compared to QG, whose physics it nevertheless depends upon in some sense.\footnote{E.g., \citet{Crowther2016, DeHaro2017, Oriti2014}, following \citet{Butterfield2011a, Butterfield2011b, Crowther2015}.} I find that this conception of emergence is likely to hold (between QG and GR, for instance), but that it does not (and cannot) play a useful role in the development and justification of the new theory.

\paragraph*{}
In this paper, I investigate the ideas of (relative) fundamentality, correspondence, reduction and emergence. Primarily, I consider these relationships as they apply between QG and GR, but I also touch upon the relationships between QG and the framework of quantum theory, and between QG and the quantum field theories of the standard model of particle physics.\footnote{Exploring the relationships between QG and quantum theory is a task that requires, and deserves, much more attention than I can devote here. There are many interesting questions regarding how quantum mechanics, and its interpretation, may be modified in the context of QG.} I have five aims in doing so:
\begin{enumerate}
\item To better understand the nature of the relationships between QG and current physics;

\item To articulate and distinguish these four types of inter-theory relations (relative fundamentality, correspondence, reduction, and emergence), and the connections between them;

\item To (begin to) expound the variety of roles that these four relations play in the context of QG, and demonstrate their utility for investigating different questions in the context of QG;

\item To encourage interest in the investigation of correspondence and reduction in the context of QG, especially in regards their roles in theory-development and justification;

\item To clarify the discussion of emergence in the QG literature, and to show how the conception of emergence applicable in QG relates to the general ideas of \textit{strong emergence} and \textit{weak emergence}, familiar from general philosophy of science.
\end{enumerate}

\paragraph*{}
The structure of the paper is as follows: \S\ref{Fund} explores what it means for a theory to be more fundamental than another, arguing that it involves an asymmetrical notion of dependence; \S\ref{FundQG} outlines how QG may be understood as more fundamental than GR. \S\ref{Corresp} presents the idea of correspondence as the broad class of inter-theory relationships intended to demonstrate that two theories whose domains of validity overlap are compatible in these domains---i.e., that the theories (approximately) share the same results in the overlap regions; this section also provides some indication of the wide variety of roles played by correspondence relations in theory development and justification. \S\ref{CorrespQG} provides examples of correspondence relations, and their roles, in the context of QG. \S\ref{Red} explores the idea of reduction, arguing that it is a special case of correspondence that applies when the region of overlap in the domains of two theories is the entire domain\footnote{I use ``domain'', ``domain of success'' and ``domain of applicability'' interchangeably to refer to the complete set of systems/phenomena successfully described by the theory in question. I take it that we are only speaking of actual, successful theories of physics, and only speaking of them as they actually apply in our world (in the case of QG, this is the theory once it is actual).} of one of these theories; but, unlike correspondence generally, it also aims at evidencing deducibility, and thus relative fundamentality. \S\ref{RedQG} shows how reduction is taken as a criterion of theory success in QG. Finally, \S\ref{Emerg} explores the conception of emergence as the novelty and robustness of a less-fundamental theory compared to a more-fundamental theory of the same system; \S\ref{EmergQG} explains how it applies in the context of QG; and \S\ref{SWeak} explores its relationship to the notions of weak emergence and strong emergence in the more general philosophy literature, arguing that while some accounts of weak emergence may be applicable, strong emergence should not hold in the case of QG/GR, given the role of reduction in the definition of the theory.

\section{Fundamentality}\label{Fund}

Here, I just speak about \textit{relative} fundamentality\footnote{While QG must be more fundamental than GR, QG need not be a fundamental theory; i.e., it is not necessary to include the criterion of (absolute) fundamentality in the definition of QG \citep{CrowtherForthcoming}.} : a \textit{more fundamental} theory, $M$, of a given system, $S$, or phenomenon, $P$, provides a \textit{more basic} description of $S$ or $P$ than a \textit{less fundamental} theory, $L$, does. There is only one condition for relative fundamentality: that the laws of $L$ somehow depend (at least partly) upon the physics described by $M$, \textit{and not vice-versa}. Note three points of clarification, however, in regards to $M$, $L$, $S$ and $P$. Firstly, $M$ will typically describe the system, $S$ at a different range of energy scales than $L$, or perhaps under different conditions. Secondly, $M$ might not actually describe the phenomenon, $P$, that $L$ does, but rather (some of) the physics underlying $P$ (i.e., part of the more-basic physics responsible for the appearance of $P$). 

\paragraph*{}
Thirdly, due to the way that theories are designated and differentiated, the more fundamental physics that is responsible for the laws of $L$, might not be described by a single more-fundamental theory (at a given scale, or level of description), but rather a set of such theories. In other words, rather than a single theory, $M$, there may be several theories, $M_\alpha$, that together provide the more-basic description of the physics of $S$ or $P$. This is what is meant by saying that the laws of $L$ depend \textit{at least partly} on the physics of $M$---it is not a claim about emergence, reduction or determinism. The idea is that the set $M_\alpha$ should be a complete description of the physics of the system at a more basic level than $L$, and this physics is responsible for that of $L$. If $M_\alpha$ comprises only a single theory, $M$, then this completely describes the physics responsible for $L$ (at a given level); if $M_\alpha$ comprises more theories than just $M$, then the physics of $L$ depends only partially upon that of $M$, and partially upon that of the other theories in the set---and the laws of $M_\alpha$ do not depend (in whole nor in part) on the physics of $L$.
 
\paragraph*{}
The apparent influence or effect of this dependence may not only be very minimal, but also obscured by the way it is incorporated into the parameters and structure of $L$. In other words, a less-fundamental physics may be largely robust and apparently autonomous, in spite of being dependent on the $M$ physics.\footnote{I maintain that we can still speak of ``dependence'', even for effective field theories that may be derived from different high-energy theories, understanding it in this minimal sense; but cf. \citet[][\S3.1]{McKenzie2017}. But, I also acknowledge, thanks to \citet{McKenzie2017}, too, that my means of characterising relative fundamentality here is likely not a unique way of doing so.} Often, the more fundamental theory provides a ``finer-grained'', or more detailed, description of the system than the less fundamental one. The $L$ is, in this sense, an approximation to $M$ that works well in a given regime (e.g., a certain range of energy scales, or under special conditions); though it is certainly not necessarily less-useful, less-accurate, nor less-apt in this regime. Also, this notion of relative fundamentality tends to correlate with higher-energy scales, or the idea that $M$ describes ``smaller stuff''. However, while these notions may be sufficient for relative fundamentality in this sense, they are not necessary.\footnote{Note again, too, that the ideas of ``higher-energy'' scales and ``shorter-length'' scales may not be useful in the context of QG (Footnote \ref{microfoot}).} 

\paragraph*{}
This account of relative fundamentality is unusual in not requiring that $M$ have a broader domain than $L$. This is so as to not rule out examples of more-fundamental theories whose domains of applicability are more limited than those of their $L$ counterparts; e.g., quantum chromodynamics (QCD) is more fundamental than atomic theory, but has a smaller domain of applicability (in terms of the range of different phenomena that it applies to, not necessarily in terms of energy scales, since QCD is UV complete). However, while the ``broader domain'' criterion is not important for defining fundamentality, it is essential for understanding reduction---and in this case we would not say that atomic theory reduces to QCD, but rather the whole of the standard model physics (\S\ref{Red}).

\subsection{Fundamentality in quantum gravity}\label{FundQG}
There are two ways in which QG is more fundamental than GR: by being quantum (or by being \textit{beyond} quantum) and by being micro\footnote{QG is necessary for describing the domains where GR and quantum field theory intersect; one of these is at extremely high energy scales.}. Each of these is sufficient to satisfy the ``asymmetric dependence'' condition of relative fundamentality. As has been well-noted in the literature, it may be that the recovery of GR from QG is a two-step process, involving both the reproduction of the ``classical appearance'', as well as the low-energy limit of the theory (which brings us back to familiar energy scales). The former is known as the quantum/classical transition, and the latter is called the micro/macro transition. While both address the question of why we do not need to use QG to describe much of the gravitational phenomena we observe, they are distinct, and may or may not be related to one another. Both transitions represent common problems in physics and the philosophy of physics; and both play a role in understanding the relationships of emergence, reduction and correspondence.

\paragraph*{}
The micro/macro transition is not something that happens to a system, but a change in the level of description---it is the ``zooming out'' to a coarser-grained theory. The micro/macro transition may be represented by an approximation procedure, a limiting process (such as the thermodynamic or ``continuum limit''), or the renormalisation group flow and the other methods of effective field theory (EFT).  The framework of EFT is employed in order to produce a theory valid at a certain energy scale from one valid at a different scale. For instance, this framework sets out a procedure for creating a low-energy (macro) theory from a high-energy (micro) one describing the same system (in terms of different degrees of freedom). It also provides an explanation for how it is that we can use our macro theories without needing to know the details of the micro-physics.\footnote{See, e.g., \citet{Batterman2005, Batterman2011, Crowther2015, Hartmann2001}.} All of these different techniques are employed by various approaches to QG,  in their attempts to connect QG back to GR.\footnote{For a discussion of the roles of approximation and limiting procedures see, e.g. \citet{Butterfield1999}, \citet{Wuthrich2017}; for a discussion of EFT and QG see \citet{Crowther2016}.}

\paragraph*{}
In the case of the quantum/classical transition there are two different issues. Quantum theories are supposed to apply universally (this fact underlies one of the main motivations for QG); so, firstly, there is the question of why, in practice, they are usually only necessary for describing small systems. Secondly, there is the \textit{measurement problem}: why it is that any measurement on a quantum system finds the system in a definite state even though the system evolves as a superposition of different states. The process of \textit{decoherence} describes how the interference effects associated with superpositions become suppressed through a system's interactions with its environment, with the consequence that the quantum nature of the system is no longer manifest. Larger systems more strongly couple to their environments, so decoherence provides the beginning of an explanation for why quantum theory is usually only necessary for describing micro-systems. As such, it gives us some insight into the ``transition'' that a system undergoes that prompts us to move from a quantum description of it to a ``classical'' one (although the system remains inherently quantum, as does the rest of the universe). Decoherence does not, however, give us an answer to the measurement problem.

\paragraph*{}                                                                                                                      We expect that the generic states of the entities described by QG will be superpositions. Yet, the quantum nature of spacetime is not manifest. A decipherment of the quantum/classical transition in QG is supposed to be necessary for understanding the classical appearance of spacetime, because limiting and approximation procedures (used to describe the micro/macro transition) alone cannot resolve a superposition.\footnote{See, e.g., the discussion in \citet{Wuthrich2017}.} Of course, this is an incredibly challenging task, given that the quantum/classical transition is poorly understood in general. It seems likely that a solution to the measurement problem is required if we are to fully understand the relationship between spacetime and the quantum degrees of freedom that somehow underlie it (or ultimately constitute it); or it may be that the solution will be provided by the theory itself.\footnote{Penrose is a firm believer in this second possibility, see e.g. \cite{Penrose1999, Penrose2002}. Also, it seems that if QG is to be a fundamental unified ``theory of everything'', then it perhaps should explain the measurement problem.} Most of the people working on QG, however, have more modest goals (for the time being, at least) and seek to better understand the relationship between QG and GR to any degree that will aid in the development, testing and justification of the theory. A rigorous philosophical understanding of the quantum/classical transition may not be required for these purposes.

\paragraph*{}
On the other hand, while QG is expected to provide a description of the physics that replaces gravity at high-energy scales, and should account for matter as well as spacetime, it is \textit{not} necessarily more fundamental than the QFTs of the standard model. One might disagree, and attempt to argue that QG is necessarily more fundamental than the standard model because these QFTs make use of a conception of spacetime, and if this spacetime is to be identified with that of GR, then it does necessarily depend on the physics of QG. If this is so, then QG is, and must be, more fundamental than our QFTs. However, the conception of spacetime that features in the standard model QFTs is not necessarily GR spacetime, and there are reasons for thinking that it is better not identified with GR spacetime (e.g., QFTs in curved spacetime are very different from the familiar standard model QFTs).\footnote{This response also applies to the objection that our QFTs and GR must be describing the same spacetime, because they both apply to our world and there is only one actual spacetime! It suggests adopting a more sophisticated (or cautious) attitude when attempting to identify structures in our theories with things in the world.}

\paragraph*{}
Instead, it seems reasonable to say that the interactions described by the standard model QFTs do not depend upon gravity; and so they may not necessarily depend on the physics that underlies gravity.\footnote{This is one reason, too, we should not insist that the problems with QFT (e.g., regarding divergences) will, or must, be solved by QG, as is frequently claimed. Cf. \citet[][\S3.8.5]{Crowther2016}.} In regimes of extreme curvature---where QG is expected to be necessary---gravity does have an effect upon quantum fields. However, these regimes are not described by the standard model QFTs, and the laws of the standard model QFTs do not necessarily depend on the physics of these regimes. Nevertheless, of course, it is possible that QG and the standard model do apply to some of the same systems, and also that the standard model QFTs do depend in some way upon the physics of QG. And, if this is the case, then we would say that QG is more fundamental than these QFTs.\footnote{This is natural if we interpret the spacetime of QFT as being an approximation to GR spacetime (applicable when curvature effects are negligible), and thus an approximation to whatever the fundamental structures described by QG are that ``underlie'' spacetime.}

\section{Correspondence}\label{Corresp}
In this section, I provide an introduction to the idea of correspondence as a class of inter-theory relations, and list some of the roles it plays in theory development and justification (which are then illustrated by examples in QG, \S\ref{CorrespQG}).\footnote{This connects with aims (1--4) listed in the Introduction.} The idea of correspondence as a class of inter-theory relations is most familiar from the common understanding of the \textit{correspondence principle}, which states, roughly, that quantum mechanics reduce to classical mechanics in the domain where the latter is successful.\footnote{The correspondence principle was famously proposed by Niels Bohr in the context of old quantum theory, yet the common understanding of the principle is most certainly not what Bohr meant by it \citep{Bokulich2014}.} Thus, as this statement of the principle demonstrates, correspondence is typically understood as what \citet{Nickles1973} calls ``reduction\textsubscript{2}'' (\textit{a.k.a} the ``physicist's sense of reduction''); as I describe in \S\ref{Red}, this notion involves a variety or approximation and limiting relations, and is aimed at demonstrating that a more-fundamental theory, $M$ subsumes the entire domain of succcess of a less-fundamental theory, $L$. But, correspondence can also be understood more broadly, and it is the broader sense that I mean here. The limiting and approximation relations characteristic of reduction\textsubscript{2} can be used in the case of \textit{any} overlap in the domains of \textit{any} two theories (i.e., not just $M$ and $L$).

\paragraph*{}
However, correspondence is not restricted to the approximation and limiting relations of reduction\textsubscript{2}: it encompasses an even wider variety of relations, as I explain shortly. Another distinguishing feature is that it is not an asymmetric relation, as reduction is, but can go in both directions between two theories. So, ``correspondence'' refers to any relationships that link two theories in the domains where both apply (i.e., provide successful descriptions), that establish (or are taken to establish, or are intended to establish) the \textit{compatibility} of the two theories within these domains---i.e., that both theories approximately (i.e., including a small degree of error) have the same results in these domains. Here, I take ``results'' to include theoretical propositions as well as observational ones, and even ``larger structures'' such as derivations and explanations.\footnote{Following \citet[][p. 79]{Butterfield2001}, but in a different context.}

\paragraph*{}
This statement of correspondence reveals the concept as tautological: Two theories that both successfully describe a given domain will necessarily be compatible within that domain, \textit{by definition}. So, correspondence relations are of the most practical interest in the cases where the success of one theory within that domain has not been demonstrated. The relationships are then articulated in order to establish that the new theory does indeed successfully describe that domain, by virtue of standing in these particular relationships to a theory whose success here has been directly and thoroughly established. As such, correspondence is a  ``shortcut to results''. 

\paragraph*{}
There is a concern, however, that correspondence relations may not be sufficient to fully and certainly establish what they are intended to (i.e., that the two theories yield all the same results in the relevant domain). Indeed, the point of correspondence is as a shortcut to results---precisely to avoid having to derive \textit{all} the results from the new theory in the old domain. Thus, there is an element of subjectivity involved in the practice of science here, in deciding when the correspondence relations articulated between two theories are enough to establish that the two theories are fully compatible in the relevant domain. Correspondence is taken to hold when the relationships demonstrated between two theories are sufficient that we (experts) are convinced that correspondence holds. To signify this, I state that correspondence is the \textit{ostensible compatibility} of the two theories.

\paragraph*{}
\citet{Post1971}, and subsequent literature, speaks of the ``generalised correspondence principle'' (GCP), with various formulations (typically intended to apply only to the relationships between $M$ and $L$ theories). My formulation of the GCP is: Any two theories whose domains of success overlap must \textit{correspond} to one another in those domains---i.e., at least one correspondence relation must obtain, sufficient that the two theories are ostensibly compatible in the overlap domains. Because we are no longer speaking just of more- and less-fundamental theories, I move to a more general notation, of $N$ for ``newer theory'' (i.e., the theory in development, or the theory whose success in a given domain is unknown), and $O$ for ``older theory'' (i.e., the established theory whose success here is known).

\paragraph*{}
There are many different types of correspondence relations; \citet{Hartmann2002}, for instance, describes seven distinct forms, in a list that is not supposed to be exhaustive. I do not reproduce them all here, but two well-known examples are: \textit{Numerical Correspondence}, where $N$ and $O$ agree on the numerical values of some quantities, and \textit{Law (or Mathematical) Correspondence}, where some laws, or other equations, from $O$ also appear in $N$. A prominent form of correspondence that features in the development of QG (but does not appear on Hartmann's (2002) list) is one that may be termed \textit{Principle Correspondence}, where $N$ and $O$ share some of the same key \textit{principles}---although these may appear in different, but analogous, forms in $O$ and $N$, as illustrated in \S\ref{CorrespQG}.

\paragraph*{}
Correspondence is typically thought of as a relation going from $N$ to $O$, meaning that the new theory should satisfy constraints presented by the old theory. But correspondence is also used from $O$ to $N$, as a means of inferring (parts of) $N$ from $O$ \citep[][207]{Radder1991}. This way of using the GCP is explicitly heuristic, and does not strictly match definitions such as numerical or law correspondence---rather, it is an exploratory, ``working'' correspondence, used to discern something of the structure of $N$ from current theories, including $O$. (\citet{Bokulich2008} also explores the important roles that this $N$ to $O$ direction of correspondence played in the development of quantum mechanics).

\paragraph*{}
By virtue of demonstrating compatibility, there are many different roles that correspondence plays in physics. These may be classified into three main types: \textit{heuristic}, \textit{justificatory}, and \textit{efficient}. These general categories correspond roughly to the three stages of science, respectively: theory \textit{construction}, theory \textit{acceptance}, and theory \textit{use}. In this paper, I only discuss the first two categories, since these are the most relevant at this stage of QG research.\footnote{The efficiency roles are about justifying the continued use of the older theory, $O$, through its relationship to $N$; or refining, or correcting $O$ through its relationship to $N$.} In \S\ref{CorrespQG}, I briefly indicate how some of these roles are employed in the search for QG. Note that this list is just meant to give some indication of the variety of different ways in which correspondence is used---it is not the goal of this paper to undertake an in-depth analysis of each of these (and whether or not they can actually be achieved by any particular correspondence relations).

\begin{enumerate}
    \item \textbf{Heuristic.} Roles of correspondence in theory construction or development of $N$. As \citet{Radder1991} states, and \citet{Bokulich2008} explores in detail, these roles may be played by correspondence from the old theory, $O$ to $N$ (rather than the more familiar ``from $N$ to $O$''). Particular links between aspects of $O$ and aspects of $N$ (the theory in development), can, for instance, serve as:
    \begin{enumerate}
        \item \textbf{Guiding principles:} Tentative guides, or aspirations, that may or may not feature in the final formulation of $N$.
        \item \textbf{Postulates:} Assumed as key features of $N$.
        \item \textbf{``Data'':} Since $O$ is successful, it can act analogously to empirical data for the new theory to be built around (this also features strongly in the ``justificatory roles'').
    \end{enumerate}
    
    \item \textbf{Justificatory.} Roles of correspondence in theory acceptance; These roles are about legitimising the new theory, $N$, by appeal to the established theory, $O$. The diverse items on this list may be appealed to either, or both, as constraints  (\#), or means of confirmation (\dag). Constraints are \textit{criteria of theory acceptance} (also referred to as ``criteria of success'', or ``definitional'' criteria), meaning that a new theory will not be accepted unless it satisfies these. Means of \textit{confirmation}, on the other hand, are non-necessary, but desirable features that serve to increase credence in the theory.
    
    \begin{enumerate}
     
        \item \textbf{Preservation of success:} $N$ must be at least as successful as $O$. Compatibility ensures that the successes of $O$ are not lost in the move to $N$; Compatibility guards against (and so correspondence relations are invoked in order to minimise) ``Kuhn-losses'', which include losses in the ability to explain certain phenomena whose authenticity continues to be recognised, losses of scientific problems (a narrowing of the field of research), an increased specialisation and increased difficulty in communicating with outsiders.\footnote{Definition from \citet[][p. 260]{Hoy1993}.} Correspondence relations may also be used to evaluate ``acceptable losses''. [\# i.e., $N$ will not be accepted unless it satisfies this criterion of ``preserving the success'' of its predecessor.]
        \item \textbf{Explanation of success:} $N$ should explain why $O$ is as successful as it is. The old theory, from the perspective of the new theory, is (to some degree) incorrect, yet it is successful by virtue of being compatible with the new theory in the relevant domain. [\#]
        \item \textbf{Identification:} Correspondence can help identify that $N$ actually describes the same systems as $O$, but at different scales (or under different conditions). However, it does not standardly demonstrate that one theory is more fundamental than another. For this, reduction is required \S\ref{Red}.  [\#]
        \item \textbf{Problem-solving explanation:} $N$ may explain, or ``explain away'', features of $O$ that are problematic, or otherwise apparently stand in need of explanation. The solution to particular such problems may be set as part of the criteria of acceptance of $N$ (i.e., in the definition of what would count as a successful theory of $N$), or be unexpected successes of $N$ that serve as additional evidence for its being correct. [\# \dag] 
        \item \textbf{Non-empirical confirmation:} As \citet{Dawid2013a} has claimed in his ``meta-inductive argument'', our credence in a theory $N$ may be increased in virtue of its standing in particular relationships to established theories and frameworks; e.g., by $N$ exemplifying some key features of $O$; empirical evidence for the other theories can indirectly also support $N$. [\dag]
        \item \textbf{``Predictions'':} Since $O$ is successful, $N$ unexpectedly recovering particular aspects of $O$ in the relevant domain, can be viewed as support for $N$. In order for these ``predictions'' to lend support for $N$, the recovered features of $O$ should be those that are actually involved in $O$'s success.\footnote{I use scare quotes because these are not actually predictions, but postdictions!} [\dag]
       \item \textbf{Local beables:} For a theory lacking conceptions of space and time (and thus ``local beables'' \citep{Bell1987}), making contact with established spatiotemporal theories is one means of deriving local beables \citep{Huggett2013}. [\# But, while the possession of local beables is a criterion of acceptance, correspondence is not a necessary means].
       \item \textbf{Empirical coherence:} Experimental testing is necessarily carried out in space and time, so the means of testing a theory that says there is no space and time may undermine the reasons for believing the theory correct; correspondence with established spatiotemporal theories is one means of avoiding this problem of ``empirical incoherence'' \citep{Huggett2013}. [\# But, while empirical coherence is a criterion of acceptance, correspondence is not a necessary means].

    \end{enumerate}
    
     
    
\end{enumerate}

\subsection{Correspondence in quantum gravity}\label{CorrespQG}
While many of these heuristic and justificatory roles are also played by novel experimental data and empirical predictions of the new theory, such data and testable predictions are notoriously lacking in the search for QG. Thus, these features of correspondence are especially significant in this context. The most obvious example of the heuristic roles is the number of approaches that take GR as a ``starting point'' for their exploration, and work forward from there (``from $O$ to $N$'')---for instance, quantised GR approaches, and other approaches based on the path-integral methods, such as causal dynamical triangulations. 

\paragraph*{}
The role of background independence in LQG is a good example of a heuristic guiding principle (1.a) being used in the development of the theory. The approach prides itself on doing justice to what its proponents recognise as one of the key principles of GR---background independence---and it is often contrasted against string theory, which seems to favour the picture of QFT over the insights of GR.\footnote{String theory is based on the covariant quantisation of gravity, which requires that the gravitational field be split into a sum of a ``background'' part, and a dynamical part (which is then quantised). This move puts the theory into the form of a QFT (although string theorists believe that the spacetime background they use in this procedure is ultimately composed of strings itself), and is rejected by LQG \citep[e.g.,][\S1.1.3--1.2.1]{Rovelli2004}.} \citet[][\S1.1.3]{Rovelli2004} describes GR as having elevated spacetime from a fixed background structure to a dynamical field, so that---figuratively speaking---there are no longer entities ``living on spacetime'', but rather a collection of fields ``living on one another''. This is (one aspect of) the general notion of background independence that LQG seeks to preserve.

\paragraph*{}
As \citet[][\S1.2.1]{Rovelli2004} goes on to explain, the implementation of a notion of background independence in LQG actually helped solve the problems with loop states---of being ``too singular'' and ``too many''. Previously, defining loop states with respect to a background structure meant that an infinitesimal change produced a new state, and thus there were a continuum of states, and no basis for a separable Hilbert space for the theory. By instead defining the loop states with respect to other loops, and having them ``live on one another'', these problems disappear, and the loop states become viable as a basis for a separable Hilbert space. The successful implementation of background independence in this way thus contributed to the progress of theory's development. The principle---having already been shown to be successful through its featuring in GR---is again vindicated. LQG, in turn, enjoys some increase in its credence, in virtue of also successfully featuring this principle in an essential way---and thus provides also an example of the ``non-empirical confirmation'' role of correspondence (2.e). 

\paragraph*{}
The correspondence relation here is what I called ``principle correspondence'' (\S\ref{Corresp}). It plays the same heuristic and justificatory roles in other approaches to QG---including causal set theory, where the key principle is Lorentz invariance. This principle is not only an essential feature of special relativity (SR) and GR---and thus well-supported given the success of these theories---but is well-confirmed independently through experiments. Any violations of it, even at the Planck scale, must be very tightly constrained so as to be indetectable.\footnote{See, e.g., \citet{Collins2009, Liberati2011}.} Lorentz invariance is thus a strong choice as a guiding principle (1.a), postulate (1.b), and ``datum'' (1.c) for QG; in causal set theory, it is used in all three of these heuristic roles. It is accomplished in the theory through the random ``sprinking'' process by which causal sets are constructed, which ensures there is no preferred frame that results. The demonstration that Lorentz invariance can be preserved even in an approach to QG that describes discrete elements is one of the biggest achievements of causal set theory \citep{Dowker2004, Dowker2005}. By featuring a key principle of previous successful theories, the implementation of Lorentz invariance in causal set theory thus serves also in the justificatory role of non-empirical confirmation (2.e). However, given the experimental evidence for Lorentz invariance, this correspondence may serve an even stronger role as support for the approach---i.e., it may count as empirical- as well as non-empirical confirmation.

\paragraph*{}
As these examples show, successfully using correspondence from $O$ to $N$ in the heuristic role of theory-construction can also serve as part of theory-justification, even if such correspondence is ``built in'' by hand, rather than derived as a ``prediction''. However, having correspondence in this way is no guarantee of having it in the other direction (from $N$ to $O$), which is the ``recovery'' direction, and the one more strongly associated with theory-justification. Both LQG and causal set theory describe structures that correspond to spacetimes (these structures are originally constructed from spacetimes, i.e., heuristic, $O$ to $N$ correspondence), yet both approaches have difficulty recovering spacetime from the multitudes of other possible structures described by their theories. They each seek a dynamics that naturally ``picks out'' the structures in their theory that correspond to spacetimes in the appropriate domains.\footnote{See, e.g., \citet{Rovelli2014,Wuthrich2017} and \citet{Dowker2005,Henson2009,Wuthrich2012,Sorkin2005}.}

\paragraph*{}
The justificatory role of ``preservation of success'' (2.a) is difficult to evaluate at this stage in QG, and is potentially only demonstrable post-hoc, once we have a close-to-fully-developed theory that is otherwise acceptable as a replacement for GR in the relevant domains. This is likely also true of ``explanation of success'' (2.b), although there are some indications of how it might work in QG. An example comes from one of the forms of correspondence in string theory: in order for the theory to be well-defined, the background spacetime containing the string must satisfy an equation that has the Einstein field equations (the central equations of GR) as a large-distance limit. If string theory replaces GR as a more-fundamental theory, then this correspondence may be appealed to in order to explain the success of GR: Even though GR is ``incorrect''---or, rather, \textit{approximate}---in some sense, it works because it features, or approximates, some aspects of string theory. In other words, the idea is that GR is successful partly in virtue of employing the Einstein field equations in the appropriate domain. The shared importance of these equations is an example of ``law correspondence'' (\S\ref{Corresp}). \citet{Huggett2015} argue that the background metric field that features in the theory is composed of stringy excitations and, given that it satisfies the Einstein field equations, is to be identified with the gravitational field. Hence, this correspondence also plays the ``identificatory role'' (2.c), which is a justification of the theory through the demonstration that it satisfies one of the criteria of QG---being that it also describes the same systems as GR (this is important in discussing reduction, \S\ref{RedQG}).

\paragraph*{}
An example of the ``problem-solving explanatory'' role (2.d), is the goal of singularity resolution---which (depending on the approach under consideration) may be set as [\#] a criterion of success of QG (i.e., a necessary requirement of the theory), or just serve as [\dag] a means of confirmation (i.e., an ``added bonus'' of an otherwise-acceptable theory, and additional evidence of its correctness). The idea is that some singularities of GR stand in need of either explanation or resolution, and some approaches to QG aim to acheive this---for instance, string theory, and loop quantum cosmology. An example of a correspondence acting as a ``prediction'' (2.f) is the recovery of the correct spacetime dimension---in accordance with SR and GR, in the appropriate domains---by CDT \citep{Ambjorn2004}. This is a form of ``numerical correspondence''; it lends support for the approach, given that the the dimension of spacetime in SR and GR is likely involved in the success of these theories and their descriptions of spacetime in the regimes where they apply.

\paragraph*{}
In regards to (2.g) and (2.h): correspondence relations may be used to solve the problems of establishing ``local beables'' and ``empirical coherence'' for a theory of QG if it does not feature any notions of space and time. ``Local beables'' is, \citet{Huggett2013} explain, a concept of John Bell, who meant it to refer to the things that we take to be real, and which are definitely associated with particular spacetime regions. The issue is that a theory without local beables is not only apparently unable to be experimentally verified (since our experiments necessarily only involve local beables), but that it may be empirically incoherent: its means of verification may undermine the reasons for believing it correct. \citet{Huggett2013}, however, show how---for a variety of QG approaches---one can potentially derive local beables, and thus avoid the challenge of empirical incoherence. In some cases, this is done by establishing correspondence relations with GR, but in others it is not necessary to make contact with full GR spacetime in order to find a notion of local beables (and so correspondence is not necessarily required for these two roles).

\paragraph*{}
Finally, I briefly mention the idea of \textit{duality}. Perhaps confusingly, duality sometimes also goes by the name of ``correspondence'' (after the AdS/CFT correspondence, or Maldecena conjecture), although it is a different concept. In the introduction to their recent special issue on the topic, \citet{Castellani2017} characterise duality as a type of symmetry, but one that relates different theories, or very different regimes of a single theory. A pair of theories is said to be \textit{dual} when they are theoretically equivalent (which may be characterised as there being an isomorphism between the models of the two theories). There are numerous examples of dualities in QG, particularly in string theory.\footnote{Please refer to the papers in the aforementioned special issue. For philosophical introductions to string dualities, see, e.g., \citet{Matsubara2013, Rickles2011}.} I cannot adequately explore these here, except to say that duality may be understood as a special case of (what I here call) correspondence---where the domains of overlap between the two theories comprises the entire domains of each (or, we could say\footnote{Thanks to an anonymous referee for suggesting this.} that the correspondence consists in an isomorphism between the physical structures of two models of the different theories).

\section{Reduction}\label{Red}
I adopt Rosaler's (2017, \S2) useful distinction between \textit{conceptions of} reduction and \textit{approaches to} reduction: a conception of reduction is a particular meaning that one attaches to the word ``reduction'', while an approach to reduction is a particular strategy for showing that some given conception of reduction holds. There are two main accounts of reduction---which can each either be understood as conceptions of reduction or approaches to reduction---the ``philosopher's account'' and the ``physicist's account'' (note that these are standardly-employed, contrastive labels only, and not intended to characterise the views of all physicists  or all philosophers). \citet{Nickles1973} calls the former \textit{reduction\textsubscript{1}}, and the latter \textit{reduction\textsubscript{2}}. Reduction\textsubscript{1} is, roughly, the idea that the laws of the less-fundamental theory, $L$, can be deduced from those of the more-fundamental theory, $M$. It is exemplified by Nagel-Schaffner reduction (after \citet{Nagel1961} and \citet{Schaffner1976}), which states that $L$ reduces to $M$ only if there is a corrected version, $L^*$, such that (Schaffner, 1976, p. 618):
\begin{enumerate}
	\item The primitive terms of $L^*$ are associated via bridge laws with various terms of $M$.
	\item$L^*$ is derivable from $M$ when it is supplemented with the bridge laws specified in 1.
	\item $L^*$ corrects $L$ in that it makes more accurate predictions than $S$ does.
	\item $L$ is explained by $M$ in that $L$ and $L^*$ are strongly analogous to one another, and $M$ indicates why $L$ works as well as it does in its domain of validity.
\end{enumerate}

Despite its difficulties, plus some objections and developments, it is fair to say that something like this sense of reduction\textsubscript{1} remains a common understanding of reduction in philosophy.\footnote{See \cite{Batterman2016, Butterfield2011a, Butterfield2011b, Dizadji2010, Riel2016}.}

\paragraph*{}
Reduction\textsubscript{2} holds that the more-fundamental theory, $M$ reduces to the less-fundamental theory, $L$ in the domains where $L$ is known to be successful. Typically, reduction\textsubscript{2} is thought to involve a characteristic limiting relation, with the reduction\textsubscript{2} of SR to non-relativistic classical mechanics in the limit of low velocities (i.e. $v/c\rightarrow 0$) often cited as a paradigmatic case. However, while there is no general account of reduction\textsubscript{2}, it is clear that it ``involves a varied collection of intertheoretic relations rather than a single, distinctive logical or mathematical relation'' (Nickles, 1973, p. 185). The whole idea is much broader than reduction\textsubscript{1}, in that it is not restricted to logical derivation (including bridge laws); also, reduction\textsubscript{2} is much weaker than reduction\textsubscript{1}, in that it is not restricted to something like the the Nagel-Schaffner schema. Additionally, it is emphasised that, in the majority of cases, it is not the entirety of theory $L$ that will be recovered by performing these operations, and neither will all the equations of $M$ feature in the reduction\textsubscript{2}; rather, $M$ and $L$ usually stand for theory \textit{parts}. As should already by clear from the discussion of correspondence, \S\ref{Corresp}, the relations of reduction\textsubscript{2} are a subset of correspondence relations---those that go from $N$ to $O$ (and perhaps also including those that have no ``directedness'').

\paragraph*{}
Rosaler (2017) takes both reduction\textsubscript{1} and reduction\textsubscript{2} as \textit{approaches to} reduction, and the \textit{conception of} reduction as \textit{domain subsumption}---meaning that both reduction\textsubscript{1} and reduction\textsubscript{2} aim at demonstrating that all the physical phenomena successfully described by theory $O$ are also, at least as successfully, described by theory $N$ (following the ``philosopher's convention'', as \citet{Rosaler2017} does, we would then say that $O$ \textit{reduces to} $N$). A quirk of this conception of reduction, however, is that one theory subsuming the domain of another neither entails nor requires that one theory is more fundamental than the other (on the above definition of fundamentality, \S\ref{Fund})---i.e., there is no associated notion of \textit{dependence}. Thus, taking reduction\textsubscript{1} and reduction\textsubscript{2} as approaches to domain subsumption, rather than conceptions of reduction in their own right, means saying that neither reduction\textsubscript{1} nor reduction\textsubscript{2} aims at demonstrating that one theory is more fundamental than another.\footnote{If relative fundamentality was defined as the $M$ having a broader domain than $L$ (rather than the asymmetrical dependence definition adopted here), then the conception of reduction as domain subsumption could aim at establishing relative fundamentality. However, this would rule out some examples of theories with relatively-restricted domains that would otherwise nevertheless be considered relatively fundamental, e.g. standard model QFTs understood as individual theories, \S\ref{Fund}.} 

\paragraph*{}
While this quirk is not necessarily a problem for the account of reduction, it does not square with the ways reduction is used in QG.\footnote{Rosaler (2017) develops a local, model-based account of reduction, and readers may be curious about the role of \textit{models} as mediating between theory and phenomena in QG. This is the field of \textit{QG phenomenology}, which may be conducted at the general level, or for specific approaches, e.g., LQG phenomenology. As yet, no philosophical work has been done on this, but see, \citet{A-C2005, Liberati2011}.} Instead, in this context, it is more useful to understand the conception of reduction as \textit{reduction\textsubscript{1}}, and to consider reduction\textsubscript{2} as an approach to evidencing the deducibility of theory $O$ from theory $N$. Both reduction\textsubscript{1} and reduction\textsubscript{2} still may be understood as aiming at domain subsumption on this account, because, if $O$ is deducible from $N$, then $N$ too describes all the phenomena that $O$ does. But, importantly, taking reduction\textsubscript{1}---instead of domain subsumption---as the conception of reduction means that these approaches can also be understood as aiming at demonstrating the relative fundamentality of $N$ compared to $O$. This is because, presumably, if $O$ can be deduced from $N$, then $O$ may be said to be \textit{dependent} upon $N$ in some sense.

\paragraph*{}
The correspondence relations, including the variety of limiting and approximation procedures, of reduction\textsubscript{2} may be interpreted as imperfect attempts toward demonstrating that $O$ is deducible from $N$---these relations can be thought of as a ``shortcut'' to, or ``stand-ins'' for, logical deduction of a corrected version of $O$ in the relevant domain from $N$. As such, again, there is an element of subjectivity  involved, in deciding when the relations have been established sufficient that we are convinced that $O$ is actually deducible from $N$ \textit{in principle}. (However, we may rightly question not only whether reduction\textsubscript{1} ever actually obtains in physics, but also the very meaningfulness of this ``in principle'' claim---as such, this aim of reduction\textsubscript{2} as an approach to reduction\textsubscript{1} may not actually be achievable). Recall that correspondence is a shortcut to demonstrating the compatibility (same results) of two theories in the domain where they both apply: domain subsumption is a special case of domain overlap, which comprises the entire domain\footnote{Again, recall that ``domain'' refers to \textit{domain of success}; i.e., the set of systems/phenomena successfully described by the theory in question.} of $O$---thus, reduction (henceforth taken to mean reduction\textsubscript{2} as an approach to reduction\textsubscript{1}) is also a special case of correspondence. It is an attempt to demonstrate that $N$ reproduces all the successful results of $O$, through the articulation of various links to $O$ in the relevant domains, intended to evidence the ``in principle deducibility'' of $O$ from $N$.

\paragraph*{}
Because reduction shares the aim of demonstrating compatibility, it plays all the justificatory roles of correspondence (2.a--g) above; but, because it also aims to demonstrate that $O$ can be deduced from $N$ in principle, it plays one additional role:

\begin{description}
  \item  \textbf{2.} (i) \textbf{Demonstrating relative fundamentality:} If $O$ reduces to $N$ (using the ``philosopher's convention'' of terminology), then the laws of $O$ depend upon the physics of $N$.   
\end{description}

\paragraph*{}

One additional comment is required in regards to reduction. The subjective aspect of judging a putative reduction (i.e., when the correspondence relations between the two theories are sufficient that we are convinced that reduction\textsubscript{1} obtains in principle) leaves room for at least two different senses of emergence, understood as a failure of reduction. Physicists might accept a theory of QG, satisfied by the strength of the correspondence relations between it and GR demonstrating that the physics it describes is responsible for gravitational phenomena. And yet, we, as philosophers, might still be unconvinced that the relations establish the ``in principle deducibility'' of GR (corrected version thereof, in relevant domain)---for instance, if a characteristic limiting relation is involved and is singular, or otherwise unrealistic. This failure of reduction in practice is the basis for an understanding of emergence; and its potentially signalling a failure in principle is yet another sense of emergence (discussed below, as strong emergence, \S\ref{SWeak}). Neither of these are what I mean by emergence here. Although they are very interesting questions, unfortunately these ideas just cannot be meaningfully explored at this stage of QG research. The appropriate conception of emergence in QG is a positive one---based on the novelty and autonomy of the emergent physics rather than a failure of reduction in any sense (\S\ref{Emerg}).

\subsection{Reduction in quantum gravity}\label{RedQG}

By definition, QG is required to be a more fundamental theory than GR, and to subsume the domain of GR. This means that the physics described by QG is supposed to be responsible for the success of the laws of GR, and that QG also describes all of the systems/phenomena that GR does. Thus, GR must reduce to QG. The correspondence relations articulated between GR and particular approaches to QG---such as those mentioned in \S\ref{CorrespQG}---are all reflections of this fact. As explained above, these relations are intended to demonstrate that QG is compatible with GR in the domains of overlap of the two theories---which is supposed to be the entire domain of GR. Reduction, on the account given here, however, means that this compatibility obtains because GR (corrected version, in relevant domain) is in principle deducible (in the sense of reduction\textsubscript{1}) from QG. This ``in principle deducibility'' is supposed to be demonstrated once the correspondence relations (of reduction\textsubscript{2}) between the two theories have been established sufficient to convince us that reduction\textsubscript{1} holds in principle (again, however, assuming the meaningfulness of such a statement). By virtue of this, these relations are able to play all of the roles listed in \S\ref{Corresp} for correspondence generally (although the heuristic roles are typically not played by the characteristic relations of reduction\textsubscript{2}, since these tend to go ``from $N$ to $O$''). 

\paragraph*{}
Additionally, some of the justificatory roles are strengthened by the relation between the two theories being one of reduction, rather than correspondence generally, because the notion of fundamentality is involved. For instance, the success of GR is explained (2.b) by its being an approximate, less-fundamental description of the more-fundamental QG physics in the appropriate regime (rather than just being compatible with the other theory in this domain). And, not only can QG and GR be said to describe (some of) the same systems (as is achieved by correspondence generally, in the ``identificatory role'' 2.c), but we can say also that the physics of QG is responsible for the laws of GR, because GR is ``in principle deducible'' from QG. The correspondence relations so-far articulated between GR and particular approaches to QG, however, are not enough to establish that GR reduces to any particular approach; indeed, reduction is likely only demonstrable post-hoc, once we have a close-to-fully-developed theory of QG.

\paragraph*{}
In regards to quantum theory: while it is possible that QG is more fundamental than the QFTs of the standard model (\S\ref{FundQG}), it is not expected (by virtue of being QG) to subsume their domain, nor for these theories to be ``derivable in principle'' from QG. Thus, reduction is inapplicable in this case (unless further criteria are added to the definition of QG).

\paragraph*{}
Before moving on, I mention the seminal work on the topic: \citet{Butterfield1999, Butterfield2001}, which explores reduction in order to understand emergence. \citet{Butterfield1999, Butterfield2001} take emergence to be a ``weaker alternative to reduction'' that they find is not captured by the formal notion of supervenience, and take reduction as an intuitive notion which they show is not captured by definitional extension, nor by accounts such as Nagel's, which rely on definitional extension. Instead, Butterfield and Isham (1999) paint a ``heterogeneous picture of emergence'' (p. 125), and ask us to ``bear in mind the variety of ways that theories can be related: in particular, with one theory being in some sense a limit of the other, or an approximation to it'' (p. 115). We can read \citet[][\S2]{Butterfield1999} as establishing that a particular \textit{approach to} reduction, being reduction\textsubscript{1}, is sometimes too weak and sometimes too strong for capturing the authors' \textit{conception of} reduction (which itself is not specified). I do not share their intuitive notion of emergence as being a weaker alternative to reduction---rather, I take it to be a notion that is resolutely distinct from, yet nevertheless stronger than, reduction, as I discuss in \S\ref{Emerg}.\footnote{In later work, Butterfield (2011a,b) argues that emergence and reduction are distinct relations, and independent of one another (rather than being stronger or weaker variants of the same type of inter-theory relation).} Instead, what \citet{Butterfield1999} takes as ``emergence'' is---I argue---more akin (if not equivalent) to \textit{reduction}\textsubscript{2}: a weaker, broader alternative to reduction\textsubscript{1}, that involves a variety of limiting and approximation procedures, and which resists a general definition because of its heterogeneous nature.

\section{Emergence}\label{Emerg}

As mentioned above (\S\ref{Intro}), the literature on ``emergence'' in QG uses the term to refer to three main things: a generic, asymmetric inter-theory relation; as something akin to reduction (indicating the ``recovery'' of GR from QG); and as a means of understanding the novelty and robustness of GR compared to QG. While these are not incompatible, they should be distinguished, as they usefully refer to distinct ideas, and play different roles in QG. The second notion, I maintain, is better understood as reduction or correspondence, rather than emergence. It is the third notion that I here advocate as the \textit{conception of} emergence in QG.\footnote{Cf. \citet{Butterfield2011a,Butterfield2011b}; for additional arguments for this conception of emergence being useful in physics, see \citet{Crowther2015,Crowther2016}.} It fits with the two general conditions that are routinely taken, in philosophy, to characterise emergence in science, which are:
\begin{description}
	\item[Dependence] Emergent phenomena are dependent on, constituted by, generated by, underlying processes.	
	\item[Independence] Emergent phenomena are autonomous from underlying processes.
\end{description}
Where the notion of ``dependence'' that features in the first condition, and the ``autonomy'' that features in the second, vary from case to case. In QG, the dependence claim is provided by the idea of \textit{relative fundamentality} above, (\S\ref{Fund}), being characterised, essentially, as asymmetrical dependence. The independence condition is comprised of two claims (each necessary): that the emergent features are \textit{novel}, and that they are \textit{robust}, compared to the more-fundamental physics.  So, a theory, $E$ \textit{emerges} from theory $M$ (or set of theories $M_\alpha$), if $M$ provides a complete more fundamental description of $S$ than $E$ does (dependence), yet $E$ is novel and robust compared to $M$ (independence). 

\paragraph*{}
Recalling the definition of relative fundamentality, the dependence claim means that $E$ depends on $M$, and not vice versa---i.e., that the laws of $E$ supervene on the physics of $M$, and that they do so in virtue of the $E$ physics being grounded in (metaphysically caused by) that of $M$. The idea of \textit{novelty} means that $E$ and $M$ are suitably distinct: there are features of $E$ that are not features of $M$, which cannot be exhibited by---or clearly expressed in terms of---the more fundamental physics. That is, there are striking differences between the two levels of description, requiring the use of different degrees of freedom. The \textit{robustness} of the emergent physics means that it is largely impervious to changes in the details of the more fundamental physics. Note that the ideas of emergence and reduction are not incompatible.\footnote{See, \citet{Butterfield2011b,Crowther2015}.} In fact, because the ``dependence'' (relative fundamentality) required for identifying cases of emergence may be established by reduction (\S\ref{Red}), the relations often go together---this is likely to be the case in QG, given the role of reduction in defining the theory. Also, the novelty and robustness aspects of emergence can sometimes be explained by reduction, through the use of the various limiting, approximation, and other correspondence relations it employs. The fact that the emergent features may be explained through reduction or correspondence relations does not in any way diminish their status as emergent entities. 

\paragraph*{}
These ideas of dependence, novelty and robustness that feature in the \textit{conception of} emergence are rough as they stand, and are to be more precisely filled-in by particular \textit{approaches to} emergence. In the case of QG, the approaches to emergence are developed by considering particular QG programs, as discussed below (\S\ref{EmergQG}). 
\paragraph*{}
Note that emergence is not the type of relation that can play a useful role in theory development. It does not establish a connection between the theories, but relies on there being one (i.e., dependence). So, even if it were taken as a criterion of acceptance, it does not allow one to ``reverse engineer''.  Additionally, emergence obtaining between theories, when one theory is known and the other is being sought, is actually a \textit{hindrance} to theory construction (or ``discovery''), thanks to the \textit{autonomy} condition of emergence. This means that the details of the underlying theory are obscured (if the emergent theory is known, but a more fundamental theory is being sought), and that the relevant aspects of the more fundamental theory for the emergent theory may be difficult to identify and/or distill (in the case where the emergent theory is sought).


\subsection{Emergence in quantum gravity}\label{EmergQG}
Unlike reduction and correspondence, emergence does not have to hold in the context of QG: it is not part of the definition of the theory, nor is it necessary (or even useful) for heuristic and justificatory purposes. Nevertheless, in many approaches to QG there are indications for GR being novel and robust compared to these theories. Together with the correspondence relations beginning their attempts to establish the legitimacy of each of these approaches as more-fundamental theories than GR, we can see (if we squint our eyes) that there is evidence that the three ingredients of emergence are present. In this section, I will mostly speak of the approaches to QG as if they satisfied the conditions for being QG, and were accepted as being QG---while recognising, of course, that none of these approaches may actually ever do so, and if one did, then it may do so in a form that is quite removed from its current state. 

\paragraph*{}
One might think that there could also be \textit{general} arguments for GR emerging from QG: the ways in which QG is (by definition) more fundamental than GR (\S\ref{FundQG}) could potentially establish the requisite novelty and autonomy of GR compared to the more-fundamental theory, QG. Such loose considerations, however, can be misleading and may not be particularly useful in the case of QG. In regards to novelty, for instance, the general idea is that ``being quantum'' and ``being micro'' descriptions of spacetime will ensure that QG is so radically distinct from GR that the novelty of the latter, compared to the exotic structures of the former, is virtually guaranteed. However, the structures described by QG need not be so alien. For instance, consider the asymptotic safety scenario \citep{Niedermaier2006,Weinberg2009}. This approach neither assumes, nor requires that the appropriate high-energy degrees of freedom are those of GR, but demonstrates the possibility that they nevertheless could be---i.e., asymptotic safety is a viable approach to QG, and it is not incompatible with this scenario that the familiar low-energy degrees of freedom of GR are retained at the high energy scales of QG. There are also semiclassical approaches to QG---the life of which demonstrate, too, that QG does not necessarily depart so radically from GR.\footnote{See: \citet{Callender2001c,Mattingly2005,Wuthrich2005}.} Thus, the \textit{novelty} that is required for emergence is not guaranteed by the minimal defining relationships between QG and GR. Comparatively, the \textit{robustness} of GR compared to QG is plausibly demonstrated on general grounds: GR is an incredibly successful theory at all known energy-scales, and the calculation of the leading quantum corrections confirm that it is largely insensitive to whatever the underlying quantum physics \citep{Burgess2004,Donoghue1994}.

\paragraph*{}
Rather than working at the general level, it is more instructive to consider emergence in the context of particular QG approaches. \citet{Huggett2013} explore the various ways in which the structures described by a number of QG approaches depart from spacetime. These authors insist that even in the most ``comparatively mild divergence from relativistic spacetime'' offered by the discrete, but still (in a sense) metrical, lattice structures of some QG approaches (e.g., those based on quantum causal history models) it is challenging to see how spacetime could be recovered  \citep[][p. 279]{Huggett2013}. More significant (though still one of the ``milder'' examples that these authors consider) is the transgression represented by causal set theory, for instance, whose fundamental structures differ from spacetime in three respects: the fundamental relation it posits cannot be readily interpreted as spatiotemporal (i.e., metrical); the theory lacks the structure to identify ``space'' (i.e., a spacelike hypersurface); and the theory appears to be non-local in a way unfamiliar to GR (p. 279). While \citet{Huggett2013} do not present an account of emergence, their focus emphasises the conceptual leaps required to bridge these non-spatiotemporal QG approaches, and GR. In doing so, they successfully demonstrate that the ``novelty'' aspect of emergence\footnote{These authors do not consider the ``dependence'' and ``autonomy'' aspects of emergence. However, in some cases, where the recovery of ``local beables'' they describe is achieved via a GR-reduction relation, then this may establish the dependence claim.} is satisfied in all of the approaches they look at---which include most of the main research programs.

\paragraph*{}
\citet{Crowther2016} develops an account of emergence in QG that is inspired by the fertile explorations of emergence in EFT (in particular, \citet{Cao1993,Batterman2011,Morrison2012}). On this account, the ``robustness'' (or ``autonomy'') of GR compared to the particular structures described by QG is demonstrated by these structures being underdetermined by spacetime. In other words, in many QG approaches, spacetime is multiply-realisable by (i.e., compatible with) distinct micro degrees of freedom, which shows that it is independent of the details of the micro-structures. The condensed matter approaches to QG furnish an analogy, where the emergent spacetime metric depends only on the symmetry and topology of the underlying theory, rather than the high-energy details of the system. These symmetries determine a universality class, and any system within this universality class will exemplify the same low-energy physics \citep[][\S5.2]{Crowther2016}. The asymptotic safety scenario is another example that features the concept of a universality class, and is concerned with identifying this rather than the specific details of the high-energy degrees of freedom: it indicates that different types of variables in this regime can lead to the same low-energy physics of GR \citep[][\S5.4]{Crowther2016}. Such analogies demonstrate how it may be possible that the theory of the micro-structures underlying gravity is underdetermined. This account of emergence is also applicable in several of the discrete approaches to QG, as well as GFT, where GR degrees of freedom may be conceived of as low-energy collective variables in a manner analogous to thermodynamics or hydrodynamics---largely independent of the underlying details of the system \citep[][\S6]{Crowther2016}. And \textit{within} approaches such as LQG and causal set theory, the actual micro-structure of spacetime may be underdetermined, i.e., spacetime may be multiply-realisable by different micro-structures described by a given theory.

\paragraph*{}
These are just brief illustrations of how the ideas of novelty and autonomy may feature in different QG approaches. The ``dependence'' aspect of emergence is embodied by the various correspondence relations that are developed as evidence of GR reducing to (whatever the particular approach to) QG. As explained above (\S\ref{Red}), reduction---unlike correspondence more generally---is able to establish that GR depends upon the proposed theory of QG (i.e., that the particular QG theory is more fundamental than GR). While, at this stage of theory-development, no such reduction has been demonstrated, there are many examples of correspondence relations in QG---some of which may eventually serve in establishing the reduction of GR to one of these theories. Some examples are above, \S\ref{CorrespQG}.

\subsection{Strong and weak emergence}\label{SWeak}
The two main accounts of emergence in general philosophy of science are weak and strong emergence, the general descriptions of which I take from \citet{Bedau1997} and \citet{Chalmers2006}. \textit{Strong emergence} is the most common notion of emergence in philosophy, and was the fixation of the ``British Emergentists'' in the 1920s. Translated into the terms of physical theories, it holds that the emergent phenomenon or behaviour arises from the micro-processes (dependence), yet there are features of it that are \textit{not deducible even in principle} from the theory of the micro-processes (independence). This idea of emergence would directly conflict with the definition of QG, which states that GR reduce to QG---meaning that GR (or, rather, models of GR in the relevant domains) be deducible in principle from QG. Thus, strong emergence should not apply in this context \textit{by definition}.

\paragraph*{}
 However---and although extremely unlikely---strong emergence could still hold regardless. This is due to the difficulty in establishing that something obtains ``in principle'', whether reduction or emergence. Recall that the correspondence relations employed in reduction\textsubscript{2} are intended as practical evidence that reduction\textsubscript{1} holds in principle: these relations are necessary ``shortcuts'' to full deduction. Above, I mentioned the unavoidable element of subjectivity involved in deciding when the shortcut relations have successfully achieved their goal of demonstrating the in principle deducibility (or compatibility, in cases of non-reduction correspondence). These relations may come arbitrarily close to exemplifying a complete reduction, but---unless a full demonstration of reduction\textsubscript{1} is indeed carried out---there will always be a gap between the ``in principle'', and what is demonstrated (or, indeed, demonstrable) in practice. It is this gap that leaves room for the possibility of strong emergence. In other words, it is possible that a weird scenario occurs where scientific consensus accepts that we have a theory of QG---entailing a considerable amount of evidence that GR reduces to it in principle---and yet someone could still mount an argument against these correspondence relations being sufficient to rule out strong emergence. 
 
\paragraph*{} 
\textit{Weak emergence} states, broadly, that a phenomenon is emergent if it arises from the micro-processes (dependence), yet it is \textit{unexpected} given the theory describing the micro-processes (independence). This general idea of ``unexpectedness'' is thus just the condition of \textit{novelty} that features in the conception of emergence in QG---being one of the two requirements for the ``independence'' claim of emergence (\S\ref{Emerg}). Because the novelty aspect of this conception is a requirement, if emergence holds in QG at all, then weak emergence will necessarily obtain as part of it. Some examples of novelty or ``unexpectedness'' of spacetime given the structures described by QG are mentioned above (\S\ref{EmergQG}), from \citet{Huggett2013}. This broad characterisation of weak emergence, by itself, however, is insufficient for satisfying the conception of emergence that best applies in the context of QG.

\paragraph*{}
I briefly mention two well-known approaches to weak emergence. One is from \citet{Bedau1997}, based on examples in complex systems and chaos theory (where systems are extremely sensitive to their external conditions).  On this account, a macro-state of a system is weakly emergent if, in spite of its dependence upon the micro-processes, it can \textit{only} be derived via simulation \citep{Bedau1997}. The slogan is ``computational irreducibility'': there is no simpler possible way of calculating the macro-state of the system than by the arduous procedure of inputting a ``contingent stream of external conditions'' into the micro-dynamical laws, and iterating the calculation of each micro-state until the sought macro-state has been found \citet[][p. 378]{Bedau1997}. Without a  sufficiently developed theory, it is not possible to evaluate whether or not this account of emergence holds in any of the particular approaches to QG. However, the correspondence relations involved in reduction may alleviate the need for simulation of this sort---if the relevant aspects of GR can be obtained from QG through various approximation and limiting relations in a satisfactory manner, then these can (in some cases) count as derivations. Thus, one would hope (at least from a practical perspective) that the particular ``reduction of GR to QG''\footnote{Understood per \S\ref{Red}, as the particular set of various correspondence relations that feature in the reduction\textsubscript{2} judged as sufficient evidence that reduction\textsubscript{1} obtains in principle.} that serves in the justification of the theory is strong enough that it functions also to render Bedau's weak emergence inapplicable in the context of QG. However, it is still possible that Bedau's weak emergence nevertheless obtains: if the correspondence relations that are taken to justify the theory do not allow for some particular aspect or result of GR to be neatly derived from QG, then this result could be weakly emergent in Bedau's sense.
 
\paragraph*{}
Another account is Wilson's \citeyearpar{Wilson2010} weak ontological emergence, where an emergent theory may be characterised by the elimination of degrees of freedom from the underlying theory. This account is certainly applicable if spacetime emerges as illustrated by the condensed matter approaches to QG, and it applies to GFT, and any other approaches where spatiotemporal degrees of freedom emerge as collective, low-energy variables, analogous to those of thermodynamics. It also may apply in the context of LQG, where degrees of freedom possessed by the spin foams are eliminated in the approximation and limiting procedures designed to resolve and/or wash-out their discrete nature and quantum properties in the recovery of spacetime. In other approaches, such as causal set theory, the applicability of Wilson's weak ontological emergence is more difficult to judge, however.

\section{Conclusion}

There is much to be gained by exploring the miscellany of inter-theory relations in physics over and above evaluating their usefulness for particular conceptions of reduction and emergence.\footnote{This sentiment accords with Batterman's recent views, e.g., \citet{Batterman2016, Batterman2017}.} I argued that the wide variety of relations broadly classifiable as ``correspondence relations'' have only one goal: evidencing the compatibility (same results) of any two theories in any domains where they both apply (i.e., attempting to establish the tautological statement that the theories each successfully describe the shared domains of success). By virtue of this one goal, correspondence relations are employed in an incredible variety of roles. I listed some of these roles in theory-development and theory-justification, with illustrations of their use in the context of QG. Among these is the familiar (but unexamined) notion that the new theory account for the success of the theory it ``replaces''.

\paragraph*{}
Correspondence relations generally, however, do not establish that one theory is more fundamental than another. I defined relative fundamentality as essentially asymmetrical dependence: the laws of the less-fundamental theory depend in some way on the physics of the more fundamental theory, and not vice versa. QG is taken, by definition, to be a more fundamental theory than GR---through ``being quantum'' and ``being micro'' descriptions of spacetime. Reduction, I argued, is a special case of correspondence that aims at illustrating both domain subsumption as well as relative fundamentality. Reduction plays an important role in defining (a successful theory of) QG. It employs various correspondence relations (in particular those involving approximations and limiting relations) in its attempt to establish that GR is in principle deducible (in a Nagel-Schaffner sense) from QG, by demonstrating that relevant aspects, or results, of GR are recoverable from QG (in the sense of Nickles' reduction\textsubscript{2}) in the relevant domains.

\paragraph*{}
I emphasised the importance of clearly distinguishing the general idea of reduction (as deducibility, derivability, or ``recovery'' of a less-fundamental theory from a more-fundamental theory) from that of emergence (as involving the two conditions of ``dependence'' and ``independence''). The literature on ``emergence'' in QG has suffered from ambiguousness between these two general ideas, and, as such, has had difficulty in engaging with the literature on emergence in philosophy of science. I presented an account of emergence in QG that accords with the general conception of emergence in philosophy of science---basically: relative fundamentality (dependence) plus novelty and autonomy (independence). While emergence, unlike reduction, is not part of the definition of QG, I argued---by considering particular approaches as well as general arguments---that it nevertheless is likely to hold between QG and GR. Yet, emergence does not play a useful role in theory development. The positive conception of emergence presented is independent of reduction; and, in the case of QG, reduction is likely to feature as part of the account of emergence, serving to demonstrate the dependence aspect. Finally, I briefly explored how the conception of emergence in QG relates to the ideas of ``weak emergence'' and ``strong emergence''. I found that while strong emergence should not hold by definition, weak emergence will necessarily hold if the conception of emergence in QG obtains. More specific accounts of emergence are difficult to evaluate in the context of particular approaches to QG at this stage of their development, and nor is it clear that doing so would be informative.

\section*{Funding}
	Funding for this research was provided by the Swiss National Science Foundation (105212 165702).
	
\section*{Acknowledgements}

Thank-you to Niels Linnemann, Keizo Matsubara, Christian W\"{u}thrich, Rasmus Jaksland, Alastair Wilson, Manus Visser, Juliusz Doboszewski, Vincent Lam, and two anonymous referees for their valuable feedback.
	
\bibliography{thesis}
\end{document}